\documentclass[acmtog, numeric]{acmart}

\usepackage{booktabs,multirow}  
\usepackage{amsfonts}           
\usepackage{amsmath}            
\usepackage{siunitx}
\usepackage{colortbl}       
\usepackage{xcolor}         

\setcopyright{none}

\usepackage[ruled]{algorithm2e} 

\SetAlFnt{\small}
\SetAlCapFnt{\small}
\SetAlCapNameFnt{\small}
\SetAlCapHSkip{0pt}

\begin{document}
\title{Vertex Features for Neural Global Illumination}

\author{Rui Su}
\affiliation{%
 \institution{School of Computer Science, Peking University}
   \city{Beijing}
   \country{China}}
 \email{susurui@pku.edu.cn}
 \orcid{0009-0004-9331-0311}

 \author{Honghao Dong}
 \affiliation{%
   \institution{School of Computer Science, Peking University}
   \city{Beijing}
   \country{China}}
 \email{cuteday@pku.edu.cn}
 \orcid{0000-0001-7247-1301}

 \author{Haojie Jin}
 \affiliation{%
   \institution{School of Computer Science, Peking University}
   \city{Beijing}
   \country{China}}
 \email{2201111642@pku.edu.cn}
 \orcid{0009-0007-7899-3603}

 \author{Yisong Chen}
 \affiliation{%
   \institution{School of Computer Science, Peking University}
   \city{Beijing}
   \country{China}}
 \email{chenyisong@pku.edu.cn}
 \orcid{0000-0002-3406-7751}

 \author{Guoping Wang}
\authornotemark[1]
 \affiliation{%
   \institution{School of Computer Science, Peking University}
   \city{Beijing}
   \country{China}}
 \email{wgp@pku.edu.cn}
 \orcid{0000-0001-7819-0076}

\author{Sheng Li}
\authornote{corresponding author. 
\\ Project website: \href{https://woaixuexisr.github.io/papers/neural-vertex-features}{https://woaixuexisr.github.io/papers/neural-vertex-features}.
}
\affiliation{%
  \institution{School of Computer Science, Peking University}
  \city{Beijing}
  \country{China}}
\email{lisheng@pku.edu.cn}
\orcid{0000-0002-8901-2184}

\authorsaddresses{}

\begin{abstract}
Recent research on learnable neural representations has been widely adopted in the field of 3D scene reconstruction and neural rendering applications. However, traditional feature grid representations often suffer from a substantial memory footprint, posing a significant bottleneck for modern parallel computing hardware. In this paper, we present neural vertex features, a generalized formulation of learnable representation for neural rendering tasks involving explicit mesh surfaces. Instead of uniformly distributing neural features throughout 3D space, our method stores learnable features directly at mesh vertices, leveraging the underlying geometry as a compact and structured representation for neural processing. This not only optimizes memory efficiency, but also improves feature representation by aligning compactly with the surface using task-specific geometric priors. Additionally, neural vertex features offer improved feature representation by compactly aligning with the surface using task-specific geometric priors. We validate our neural representation across diverse neural rendering tasks, with a specific emphasis on neural radiosity. Experimental results demonstrate that our method reduces memory consumption to only one-fifth (or even less) of grid-based representations, while maintaining comparable rendering quality and lowering inference overhead. 

\end{abstract}



\keywords{Neural Rendering, Encodings, Global Illumination, Radiosity}

\begin{teaserfigure}
    \centering
  \includegraphics[width=\textwidth]{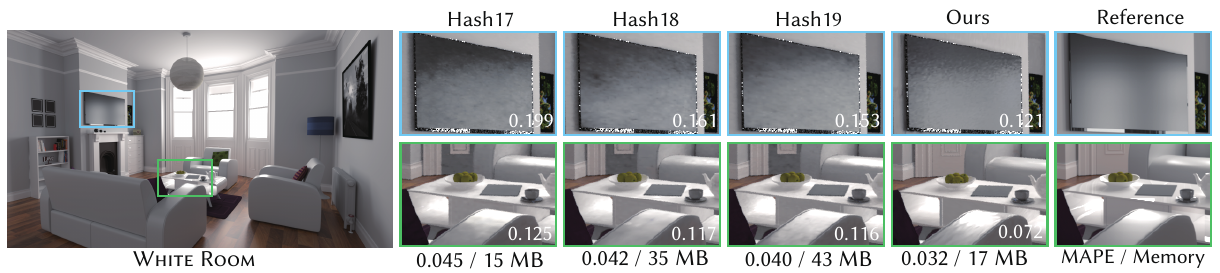}
  \vspace{-0.1in}
  \caption{
\textit{Neural vertex features} (Ours) leverages the geometric priors in neural rendering applications, effectively culling the redundant features by focusing on the occupancy of the scenes. Compared to the multi-resolution hashing encoding \cite{muller2022instant} at different sizes of hash tables ($2^{17}$-$2^{19}$ features each level), our method achieves better scalability and significantly less memory overhead at the same level of visual quality. The better compactness of our method also results in faster inference for being more friendly to the cache hierarchies of modern GPUs. }
  \label{fig:teaser}
\end{teaserfigure}

\maketitle

\section{Introduction}
\label{sec:introduction}

Global illumination has become an important topic of research in computer graphics, while recent research has leveraged  the power of deep learning to achieve realistic global illumination effects. Various methods have been developed to tackle the challenge, such as neural rendering \cite{muller2021real, hadadan2021neural, su2024dynamic, coomans2024real}. These techniques generally follow a pipeline where spatial features are extracted through input encoding and then processed by multi-layer perceptrons (MLPs) to generate the desired outputs. While these approaches have shown remarkable results in producing high-quality images, they often face significant challenges with respect to memory footprint and thus inference overhead. On the other hand, as the parallel computing hardware with tensor arithmetic accelerators iterates, more and more neural workloads tend to become memory-limited where the memory access of neural networks becomes one of the main performance bottlenecks. For example, even the cutting-edge GPUs have an L2 memory of less than 72MB, making commonly used neural networks difficult to fit into caches. 

One of the key aspects influencing the performance of neural rendering models is the input encoding. A widely adopted technique is grid-based encoding, such as the multi-resolution hash grid \cite{muller2022instant}, which scatters spatial features over structured 3D grids to enable the use of compact MLPs and accelerate convergence. While this approach is memory-efficient compared to fully dense grids due to its hash-based storage, it still allocates features in empty space rather than strictly along the scene surfaces. Even with hash-based storage, in practice, the features used in the grid account for only 55\% to 72\% of the total hash table capacity, leading to memory inefficiency. This discrepancy may cause unnecessary memory consumption, particularly in real-time applications.

Our method encodes spatial information directly at the vertices of the scene's mesh, leveraging the known geometry as a priori in neural rendering applications, and thus allowing for a more compact and memory-efficient representation of the scene. Unlike multi-resolution grids, which allocate resources to unnecessary space, our approach concentrates computation on the actual scene geometry, significantly reducing memory overhead. Furthermore, we introduce a multi-resolution representation of surface feature encoding that can dynamically refine the mesh-based feature on training loss statistics. This enables our method to capture high-frequency details even in regions that initially have low mesh density.

We evaluate our approach with diverse neural rendering applications like Neural Radiosity \cite{hadadan2021neural} and dynamic neural radiosity (DNR) \cite{su2024dynamic} using both static and dynamic scenes, as well as neural path guiding \cite{dong2023neural}. Our method demonstrates a significant improvement in memory efficiency, achieving over $5\times$ reduction in memory overhead while maintaining comparable image quality to state-of-the-art methods. 
By focusing on encoding spatial information directly at the mesh vertices and utilizing adaptive surface subdividing, we address the challenges of high memory consumption and inefficient feature representation in prior approaches. 

Overall, our main contributions are as follows:
\begin{itemize}
    \item We present a novel encoding method, neural vertex features, that stores spatial information at the vertices of the scene's mesh, offering a memory-efficient and computationally effective solution for neural rendering.
    \item We propose a multi-resolution surface learnable feature that dynamically refines the mesh based on training loss, allowing high-frequency illumination details to be accurately captured even in coarse regions with initially low resolution.
    \item Several neural rendering approaches validate the effectiveness of our formulation in significantly reducing memory usage while still achieving high-quality rendering results. This advancement may benefit a wide range of neural rendering tasks that previously depended on hash grids, offering enhanced applicability and flexibility.

\end{itemize}
\section{Related Work}
\label{sec:related}

\paragraph{Neural Rendering}

Deep learning techniques have been widely applied to facilitate rendering applications. Many works focus on image-space techniques, employing deep generative networks for tasks such as denoising \cite{vogels2018denoising, icsik2021interactive}, up-scaling \cite{wu2023extrass, zhong2023fusesr}, and directly operating in the screen space \cite{granskog2020compositional, diolatzis2022active}. Other works explore object-space techniques, such as \cite{zheng2023nelt, zheng2024neural}, where networks are used to model the light transfer function between individual objects and the environment, subsequently composing them to produce globally illuminated images.

The success of neural scene representations, particularly Neural Radiance Fields (NeRF) \cite{mildenhall2021nerf}, has further advanced the application of deep learning in photorealistic rendering. These works operate in scene-space, where neural representations model the spatial features of the scene, decoded using lightweight MLPs. Some works \cite{muller2021real, hadadan2021neural, coomans2024real, su2024dynamic} have applied these methods for efficient caching and querying of spatial radiance, while another application focuses on enhancing path tracing by encoding target distributions for path guiding \cite{dong2023neural, huang2024online}. In this work, we introduce a novel method for encoding spatial features more efficiently, which can further improve these approaches.

\paragraph{Network Encoding}
Research has also found that input encoding plays a crucial role in network performance. Early methods, such as \cite{mildenhall2021nerf}, use frequency encoding to project the 3D coordinate inputs into a high-dimensional space using Fourier basis. Another set of approaches employs trainable encoding with multi-resolution spatial grids \cite{hadadan2021neural} or hash tables \cite{muller2022instant}. The trainable encoding technique reduces the burden on pure MLP methods, resulting in lower computational costs by keeping the MLP size sufficiently small \cite{takikawa2023compact}. However, the feature grid approach is inefficient as it allocates many features to empty space, while the scene surface scales in 2D.

To address this inefficiency, feature-decomposition-based works such as \cite{fridovich2023k, cao2023hexplane, shao2023tensor4d} project the dense grid along one or more axes and combine the resulting lower-dimensional features, namely \textit{k-planes}. While these techniques ensure fewer parameters, they make the strong assumption that sparsity in the data can be well explained by axis-aligned projections, leaving the MLP to model the correlation of high dimensionalities. Another approach is learning the indexing function \cite{li2023compressing, takikawa2022variable}, where an index codebook holds the lookup indices into the feature codebook. \citet{takikawa2023compact} combine learned indexing with spatial hashing, arithmetically combining their indices. Additionally, some methods directly generate meshes from neural implicit fields and store features on them, yielding a compact representation that supports both reconstruction and deformation \cite{neumesh, mahajan2024meshFeat}.

If the surface of interest is known a priori, compact structures such as octrees \cite{takikawa2021neural} or sparse grids \cite{chabra2020deep, peng2020convolutional} can be used to cull away unused features. However, these methods primarily work on neural SDF representations for a single object and are not scalable for rendering complex indoor scenes. In this work, we propose a neural vertex feature representation that captures scene spatial features better and is better suited for rendering tasks.

\section{Background}
\label{sec:background}

\subsection{Rendering Equation}

The fundamental formulation for light transport simulation is the rendering equation \cite{kajiya1986rendering}, which expresses the outgoing radiance at a point $\mathbf{x}$ in a given direction $\omega_o$ as the sum of emitted radiance and the integral of incoming radiance, weighted by the bidirectional reflectance distribution function (BRDF):
\begin{equation}
L(\mathbf{x}, \omega_o) = L_e(\mathbf{x}, \omega_o) + \int_{\mathcal{H}^2} f_r(\mathbf{x}, \omega_i, \omega_o) L_i(\mathbf{x}, \omega_i) \left| (\mathbf{n} \cdot \omega_i) \right| \, \mathrm{d}\omega_i.
\label{eq:re}
\end{equation}

Here, $L(\mathbf{x}, \omega_o)$ denotes the outgoing radiance at point $\mathbf{x}$ in direction $\omega_o$, $L_e(\mathbf{x}, \omega_o)$ is the emitted radiance, and the integral accumulates the incoming radiance $L_i(\mathbf{x}, \omega_i)$ over the hemisphere $\mathcal{H}^2$. The term $f_r(\mathbf{x}, \omega_i, \omega_o)$ is the BRDF that describes the reflectance properties of the surface. The equation is recursive in nature, as the incoming radiance at each point is the outgoing radiance from other points in the scene, making it computationally expensive to solve.

\subsection{Neural Radiosity}

To address the complexity of solving the rendering equation, \citet{hadadan2021neural} introduced neural radiosity, using a neural network with parameters $\theta$ to predict outgoing radiance $L(\mathbf{x}, \omega_o)$. The network is optimized by minimizing the residual between the left-hand side (LHS) and the right-hand side (RHS) of the rendering equation:
\begin{multline}
r_\theta(\mathbf{x}, \omega_o) = L_\theta(\mathbf{x}, \omega_o) - L_e(\mathbf{x}, \omega_o) \\- \int_{\mathcal{H}^2} f_r(\mathbf{x}, \omega_i, \omega_o) L_\theta(\mathbf{x'}(\mathbf{x}, \omega_i), -\omega_i) \left| (\mathbf{n} \cdot \omega_i) \right| \, \mathrm{d}\omega_i,
\label{eq:residual}
\end{multline}
where $\mathbf{x'}(\mathbf{x}, \omega_i)$ denotes the closest surface intersection of the ray with origin $\mathbf{x}$ and direction $\omega_i$. The reflected radiance is computed by tracing rays in the reverse direction (RHS rays), and Monte Carlo integration is used to estimate the integral of reflected radiance. The radiance value is predicted by the neural network $L_\theta(\mathbf{x}, \omega_o)$.

This optimization is achieved by minimizing the mean squared error (MSE) loss across multiple points and directions in the scene, as described by the following objective:
\begin{equation}
\theta^* = \arg\min_\theta \int_\mathcal{S} \int_{\mathcal{H}^2} r_\theta(\mathbf{x}, \omega_o)^2 \, \mathrm{d}\omega_o \, \mathrm{d}\mathbf{x},
\label{eq:loss}
\end{equation}
where $\mathcal{S}$ represents the scene surface and $\mathcal{H}^2$ is the hemisphere over each shading point $\mathbf{x}$.

Neural Radiosity effectively leverages the radiometric prior to optimize network parameters and model the global radiance field with lightweight neural networks and trainable spatial feature grids. However, they rely on using multi-resolution feature grids for spatial encoding, which suffers from inefficiencies in memory usage, as large portions of the grid are allocated to empty space. To address these limitations, we propose a more memory-efficient approach by directly encoding neural features at the vertices of the scene's mesh. This eliminates the need for a multi-resolution grid and significantly reduces memory overhead.

\section{Method}
\label{sec:method}

In this section, we introduce neural vertex features and a multi-resolution surface representation to model spatially varying features on the scene manifold~$\mathcal{S}$. We then describe our optimization strategy and demonstrate the effectiveness of our method across various neural rendering applications.

\subsection{Neural Vertex Features}
\label{neural_vertex_features}

In neural rendering tasks, network queries are typically performed on the surface of the scene, where the model needs to retrieve and process spatial information to generate realistic lighting, shadows, and textures. Since these queries are inherently tied to the surface, it becomes natural to maintain neural features directly on the scene's surface. The explicit mesh representation of the scene, which already defines the geometry of the surface in terms of vertices and edges, naturally lends itself to this approach.

\begin{figure*}[t]
  \centering
  \includegraphics[width=0.8\linewidth]{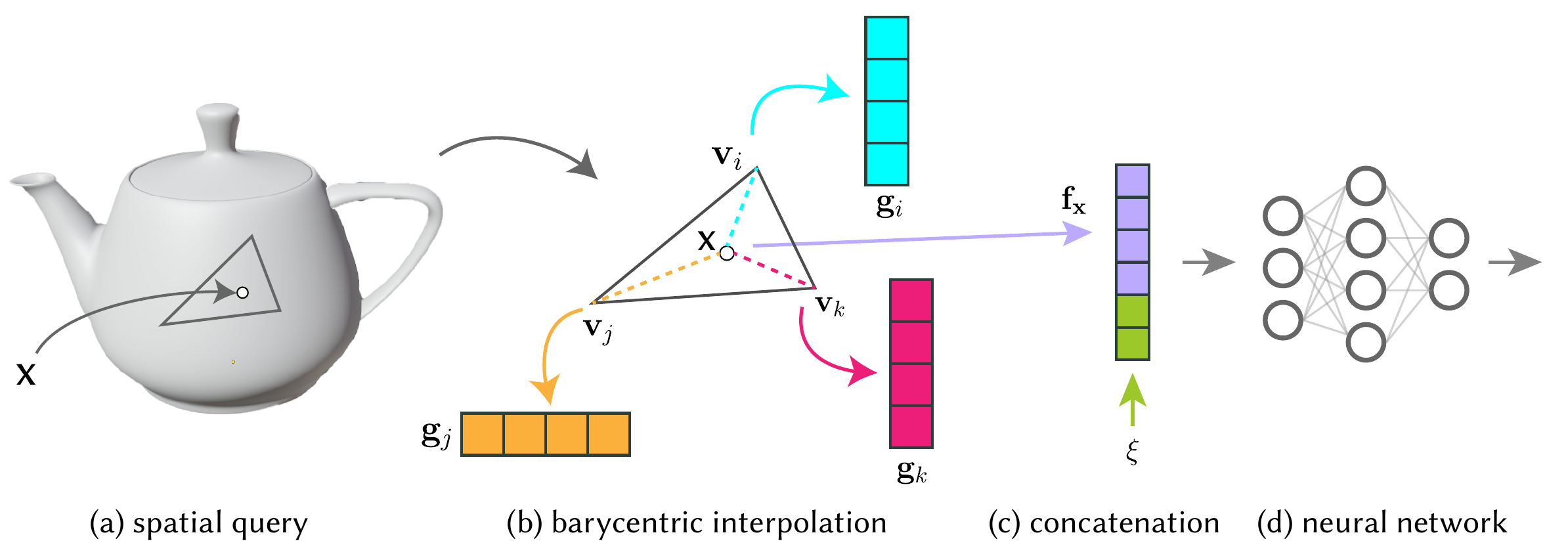}
  \vspace{-0.15in}
  \caption{Schematic of our method. The trainable features are directly attached to the geometry vertices within the scene. To support efficient modeling and querying for arbitrary positions within the domain of scene surfaces $\mathcal{S}$, we use the barycentric coordinate $(u, v)$ of the position $\mathbf{x}$ over the triangle primitive $\mathcal{P}$ to interpolate with the features in each vertex of $\mathcal{P}$, where all these features could be easily acquired in any programmable ray-tracing or rasterization pipelines. The obtained weighted feature $\mathbf{f}_\mathbf{x}$ is then used as network inputs and could be concatenated with other inputs $\xi$ to work with different downstream applications, e.g., neural radiosity~\cite{hadadan2021neural}. Implementation details are provided in Sec.~\ref{sec:implementation}.}
  \label{fig:model}
\end{figure*}

To this end, we propose neural vertex features, where neural features are stored along with each vertex of the scene's geometries. For simplicity in our experiments, we assume that the geometry consists solely of triangular faces while discussing the potential applications of alternatives, e.g., quadrilaterals. Formally, the union of all scene vertices is represented as $\mathcal{V}$ (including the virtual vertices, which will be discussed later). We maintain a feature vector array $\mathcal{G} \in \mathbb{R}^{n \times d}$, where $n = \left| \mathcal{V} \right|$ is the number of vertices and $d$ is the length of the feature vector. Each vertex $\mathbf{v}_i \in \mathcal{V}$ is uniquely mapped to a corresponding feature vector $\mathbf{g}_i \in \mathcal{G}$ based on its vertex index $i$. 

As illustrated in Fig. \ref{fig:model}, for each spatial query $\mathbf{x}$, we first identify the vertices $\{ \mathbf{v}_i, \mathbf{v}_j, \mathbf{v}_k \}$ that corresponds to the surface primitive $\mathcal{P}$. Then, we perform barycentric interpolation on the neural features associated with the vertices of the triangle primitive to obtain the feature at the queried point. 
\begin{equation}
\mathbf{f}_\mathbf{x} = (1 - u - v) \cdot \mathbf{g}_i + u \cdot \mathbf{g}_j + v \cdot \mathbf{g}_k,
\end{equation}
where $\mathbf{f}_\mathbf{x}$ is the encoded feature at the spatial point $\mathbf{x}$, and $u, v$ are the barycentric coordinates. The neural features $\{ \mathbf{g}_i, \mathbf{g}_j, \mathbf{g}_k \}$ correspond to the vertices $\{ \mathbf{v}_i, \mathbf{v}_j, \mathbf{v}_k \}$ of the triangle.

Identification of the corresponding primitive ID and calculation of the barycentric coordinates is straightforward in current rendering frameworks, whether using ray tracing or rasterization, as these operations are commonly performed during the intersection tests or fragment shading stages, with hardly any performance overhead.

\paragraph{Discussion} Neural vertex features can encode spatial information directly at the vertices of the mesh. Compared to the multi-resolution feature grid, storing neural features at the vertices enables a more memory-efficient approach. This strategy exploits the geometric priors to allow for a compact and precise representation of the scene, efficiently capturing important surface details while minimizing unnecessary storage overhead.

\subsection{Multi-resolution Surface Feature Encoding}
\label{adaptive_surface_subdivision}

While neural vertex features provide a compact and surface-aligned representation for neural rendering, their effectiveness can be limited by the density of the underlying mesh. In regions where the mesh is coarse or lacks sufficient geometric resolution, the sparse distribution of vertices may hinder the representation of high-frequency illumination details, leading to artifacts or degraded rendering quality, as shown in Fig.~\ref{fig:no_subdivide}.

\begin{figure}[h]
  \centering
  \includegraphics[width=0.9\linewidth]{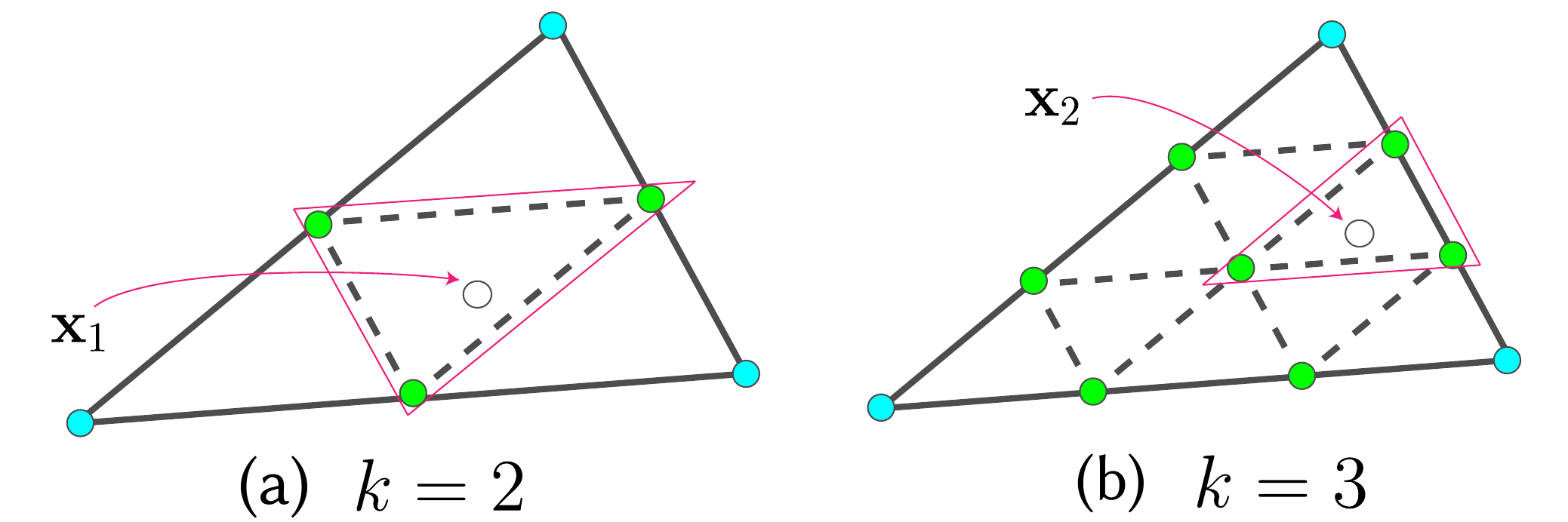}
  \vspace{-0.1in}
  \caption{Illustration of our multi-resolution feature encoding strategy. Blue vertices denote original mesh vertices, green vertices are virtual ones, and dashed lines indicate the associated LOD of features.  (a) shows $k=2$, and (b) shows $k=3$. Query points $\mathbf{x}_1$ and $\mathbf{x}_2$ lie within the subdivided feature representation highlighted by red boxes, enabling denser and more adaptive illumination feature without actually modifying the original mesh.}

  \label{fig:subdivision}
\end{figure}

To address this issue, we propose a multi-resolution strategy that increases the density of learnable features in necessary regions without modifying the actual mesh topology.  Specifically, we introduce a per-face level of detail (LOD) factor $k_i$, where $i \in [1, |\mathcal{F}|]$ indexes original triangle faces in the mesh, and $\mathcal{F}$ denotes the set of all faces.  The factor $k_i$ specifies that each edge of triangle $i$ is divided into $k_i$ segments, forming a regular triangular grid within the face, as illustrated in Fig.~\ref{fig:subdivision}.  For each resulting virtual vertex, we assign a trainable neural feature, analogous to those defined on the actual mesh vertices.

For each spatial query point $\mathbf{x}$ within an original triangle face $i$, with barycentric coordinates $(u, v)$, we determine the virtual sub-triangle in which $\mathbf{x}$ resides, and compute the corresponding local barycentric coordinates $(u', v')$ within that virtual triangle as:
\begin{equation}
\begin{aligned}
\bar{u} &= \lfloor k u \rfloor, \quad \bar{v} = \lfloor k v \rfloor, \\
\tilde{u} &= k u - \bar{u}, \quad \tilde{v} = k v - \bar{v}, 
\end{aligned}
\end{equation}

\begin{equation}
\begin{aligned}
(u', v') &= 
\begin{cases}
(\tilde{u}, \tilde{v}) & \text{if } \tilde{u} + \tilde{v} \leq 1, \\
(1 - \tilde{u}, 1 - \tilde{v}) & \text{otherwise}.
\end{cases}
\end{aligned}
\end{equation}

The virtual sub-triangle that contains $\mathbf{x}$ is identified by the integer grid coordinates $(\bar{u}, \bar{v})$. After this step, we perform the aforementioned barycentric interpolation using the features of the three corresponding virtual vertices.

All LOD factors are initially set to $k_i = 1$, meaning no virtual vertices are added.  During training, these factors are adaptively increased in response to the training loss, allocating higher spatial resolution to more challenging regions.  When a face is refined, new virtual vertices are generated, and their features are initialized via interpolation from the existing representation.  These features replace the originals, allowing the network to refine its capacity where needed. The loss-based update strategy is described in the next section.

\subsection{Optimization}
\label{optimization}

To optimize the model, we assume that the loss is defined on the scene surface. The per-point loss at $\mathbf{x}$ is denoted as $l_\theta(\mathbf{x})$, where $\theta$ represents the parameters of the network. The total loss function of the network is defined as
\begin{equation}
\label{eq:point_loss}
\mathcal{L}(\theta) = \int_\mathcal{S} l_\theta(\mathbf{x})\, \mathrm{d}\mathbf{x},
\end{equation}
where $\mathcal{S}$ represents the union of all surfaces in the scene.

To estimate the total loss, we use Monte Carlo integration. Previous work \cite{hadadan2021neural, su2024dynamic} typically employs a probability distribution proportional to the surface area to sample positions on the scene surface. However, in our model, the features are not uniformly distributed across the scene surface, leading to inefficient training. Specifically, areas with lower feature density tend to converge faster than regions with higher feature density, potentially slowing down overall optimization. To address this issue, we adopt an importance sampling strategy that initializes the sampling density based on the current feature distribution and dynamically adjusts it in sync with the evolving LOD levels.

We define the sampling probability at the triangle level, incorporating both area-based and vertex-based strategies. Specifically, let \( p(i) \) denote the sampling probability of triangle \( i \), where \( i \in [1, |\mathcal{F}|] \) indexes the original mesh faces. The area-based probability \( p_{\mathrm{area}}(i) \) and the vertex-based probability \( p_{\mathrm{vertex}}(i) \) are computed as:
\begin{equation}
p_{\mathrm{area}}(i) = \frac{A(i)}{A_{\mathrm{total}}}, \quad p_{\mathrm{vertex}}(i) = \frac{N(i)}{N_{\mathrm{total}}},
\end{equation}
where \( A(i) \) is the area of triangle \( i \), and \( N(i) \) is the number of vertices (including both real and virtual vertices) associated with the subdivided mesh \( M_i \), computed as \( N(i) = \frac{(k_i + 1)(k_i + 2)}{2} \).  
\( A_{\mathrm{total}} \) and \( N_{\mathrm{total}} \) denote the total surface area and the total number of vertices across the entire scene, respectively.

The sampling probability \( p(i) \) for triangle \( i \) is defined as a weighted combination of the area-based and vertex-based probabilities, controlled by a factor \( \alpha \):
\begin{equation}
\label{eq:prob_mix}
p(i) = \alpha \cdot p_{\mathrm{area}}(i) + (1 - \alpha) \cdot p_{\mathrm{vertex}}(i).
\end{equation}

During training, we adaptively update the LOD factor \( k_i \) to refine regions that require higher resolution. Following prior works \cite{zhao2015stochastic, diolatzis2022active}, we use the training loss statistics to guide this update. Specifically, we compute the average loss \( L(i) \) for each triangle \( i \) over a training interval and adjust the LOD factor accordingly.

Let \( L_\mathrm{mean}^\prime \) and \( L_\mathrm{std}^\prime \) denote the mean and standard deviation of the adjusted loss values \( L(i)' \), where
\begin{equation}
L(i)' = L(i) \cdot \log(R(i) + 1),
\end{equation}
with \( L(i) \) and \( R(i) \) representing the loss and radiance of triangle \( i \), respectively. The LOD factor is then updated as
\begin{equation}
k_i' = k_i + \left\lfloor \frac{L(i)' - L_\mathrm{mean}^\prime}{L_\mathrm{std}^\prime} \right\rfloor, \quad \text{if } L(i)' > L_\mathrm{mean}^\prime + 2 \cdot L_\mathrm{std}^\prime \ ,
\end{equation}
and remains unchanged otherwise.

\subsection{Rendering Task-specific Formulation}
\label{application}

For different tasks, it is necessary to select an appropriate per-point loss (Eq. \ref{eq:point_loss}). In this work, we apply our method to a neural rendering task, where the network directly outputs the radiance \( L_\theta(\mathbf{x}, \omega_o) \) at a given point \( \mathbf{x} \) in the scene and for a given outgoing direction \( \omega_o \). Specifically, we follow the Neural Radiosity method \cite{hadadan2021neural}, where the residual of the rendering equation (Eq. \ref{eq:residual}) is used as the training target to optimize the model.

\paragraph{Static Scene} On static scenes, the per-point loss is the residual of the render equation at point $\mathbf{x}$, which integrates on the hemisphere for $\omega_o$.
\begin{equation}
l_\theta(\mathbf{x})=\int_{\mathcal{H}^2}r_\theta(\mathbf{x}, \omega_o)\,\mathrm{d}\omega_o,    
\end{equation}
where the $r_\theta(\mathbf{x}, \omega_o)$ is the residual of the rendering equation and $\mathcal{H}^2$ is the hemisphere space.

Evaluation of the per-point loss is performed using Monte Carlo integration. Specifically, we uniformly sample the outgoing direction $\omega_o^j$ within the upper hemisphere of point $\mathbf{x}$. The network is then queried to obtain the radiance $L_\theta(\mathbf{x}, \omega_o^j)$. For the right-hand side of the rendering equation, we sample the reflection direction $\omega_i^{j,k}$ and intersect it with the scene at $\mathbf{x}^\prime(\mathbf{x}, \omega_i^{j,k})$. The network is queried again to obtain the radiance $L_\theta(\mathbf{x}'(\mathbf{x}, \omega_i^{j,k}), -\omega_i^{j, k})$. This process is repeated for $M$ samples, and the average value is taken as the result:
\begin{multline}
r_{\theta, MC}(\mathbf{x}, \omega_o^j) = L_\theta(\mathbf{x}, \omega_o^j) - L_e(\mathbf{x}, \omega_o^j) \\
-\frac{1}{M} \sum_{k=1}^{M} \frac{f_r(\mathbf{x}, \omega_o^j, \omega_i^{j, k}) L_\theta(\mathbf{x}'(\mathbf{x}, \omega_i^{j, k}), -\omega_i^{j, k}) \left| \mathbf{n} \cdot \omega_i^{j, k} \right|}{p(\omega_i^{j, k})},
\label{eq:our_mc_residual}
\end{multline}
where $p(\omega_i^{j,k})$ is the sampling weight according to the BSDF importance sampling.

\paragraph{Dynamic Scene} For dynamic scenes, we follow the scene definition as described in \cite{su2024dynamic}. The scene consists of $n$ animated components, each of which is predefined with starting and ending states. A scalar value $v_i \in [0, 1]$ is used to interpolate the states for each component. The scene state can be described by these values, referred to as the \textit{explicit scene vector} $\mathbf{v} \in \mathbb{R}^n$. The per-point loss is also an integral over the entire animation space $\mathcal{V}$.
\begin{figure}[h]
  \centering
  \includegraphics[width=0.9\linewidth]{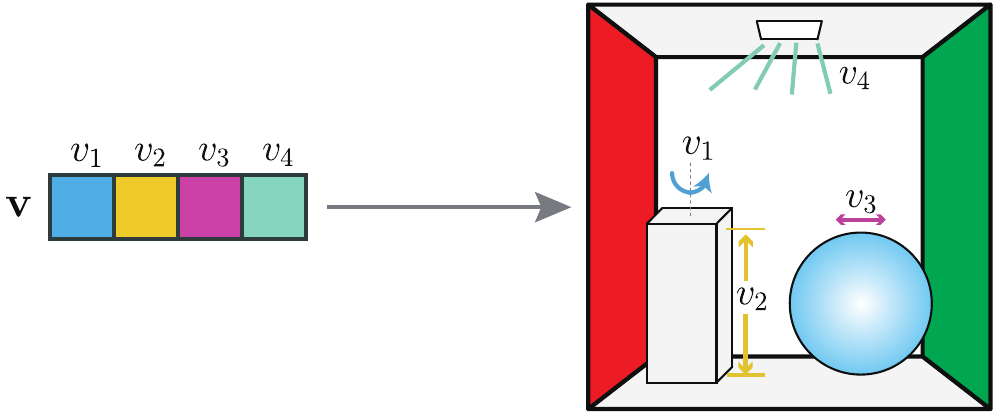}
  \caption{Illustration of dynamic scene representation. To flexibly represent a dynamic scene, we use a scalar to encode the \textit{state} of each animated component/node. The encoded domains could include the deformation ($v_2$), rigid body transformation ($v_1$, $v_3$), and lighting condition ($v_4$), for example. Together, these scalar values $v_i$ constitute the \textit{explicit scene state} $\mathbf{v}$.}
  \label{fig:dynamic_illustrate}
\end{figure}
\begin{equation}
l_\theta(\mathbf{x}) = \int_\mathcal{V} \int_{\mathcal{H}^2} r_\theta(\mathbf{x}, \omega_o, \mathbf{v}) \, \mathrm{d}\omega_o \, \mathrm{d}\mathbf{v},
\end{equation}
where $r_\theta(\mathbf{x}, \omega_o, \mathbf{v})$ is the residual of the rendering equation conditioned on the scene state $\mathbf{v}$. The evaluation of the per-point loss is similar to the static scene, with the additional step of sampling over the scene state $\mathbf{v}$.

The model should also change slightly, and we should model the radiance $L_\theta(\mathbf{x},\omega_o, \mathbf{v})$ conditioned on the scene state $\mathbf{v}$. To model the extra dimension of the animation, we follow the practice of previous work \cite{su2024dynamic}, which adds multiple feature planes for the scene vector and spatial position, along with a lightweight MLP to capture the correlations between the scene variables.
\begin{equation}
f_{\mathcal{XV}}(\mathbf{x}, \mathbf{v}) = \bigoplus_{p \in \{x, y, z\}, v \in \mathbf{v}} P(p, v),
\end{equation}
\begin{equation}
f_{\mathcal{V}}(\mathbf{v}) = \mathrm{MLP}(\texttt{Freq}(\mathbf{v})),
\end{equation}
where $P(\cdot, \cdot)$ denotes the feature plane for modeling the correlation between the variables and spatial features, while $\texttt{Freq}(\cdot)$ denotes the frequency encoding \cite{mildenhall2021nerf} applied to the scene vector $\mathbf{v}$ before feeding it into the MLP.

These two parts are added to assist with our spatial neural vertex feature encoding, which models the full dynamic radiance field $L_\theta(\mathbf{x}, \omega_o, \mathbf{v})$ effectively. The use of our neural vertex feature encoding significantly reduces memory usage while capturing the full dynamic radiance field $L_\theta(\mathbf{x}, \omega_o, \mathbf{v})$ effectively.
\section{Implementation}
\label{sec:implementation}

\paragraph{System}
We implement our algorithm using the Mitsuba3 renderer \cite{jakob2022mitsuba3}. The neural vertex features are implemented with PyTorch \cite{Paszke_PyTorch_An_Imperative_2019}, and the multi-layer perceptron (MLP) is implemented using \texttt{tiny-cuda-nn} \cite{Muller_tiny-cuda-nn_2021}. We train our model and render images on an Nvidia RTX 4090 GPU.

\paragraph{Render integration}
To render dynamic scenes with our proposed method, we first ray-intersect the camera rays to obtain the primary shading point, along with the view directions $\omega_o$ and geometry information. This data is then fed into the network to obtain the approximated radiance $L_\theta$. In our experiments, we perform offline rendering with 32 rays per pixel for static scenes, and for interactive rendering of dynamic scenes, we use 1 ray per pixel with FXAA \cite{fxaa_} assistance.

\begin{figure*}[t]
  \centering
  \includegraphics[width=0.95\linewidth]{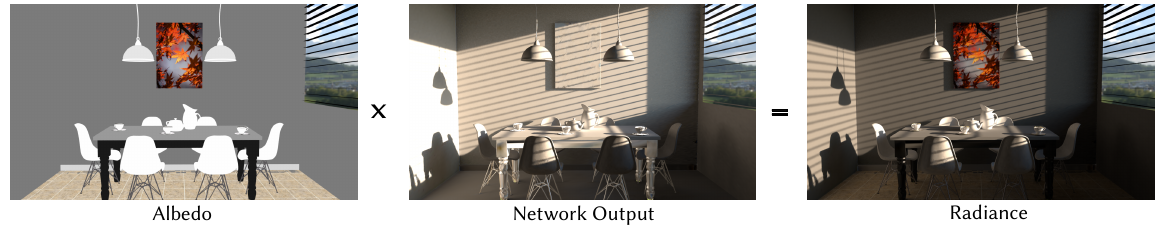}
\caption{Visualization of reflectance factorization. We let the network predict the direction-independent irradiance term by factorizing out the reflectance term. This technique helps to alleviate the burden of the network while helping recover the high-frequency details of the textures.}  
  \label{fig:reflectance}
\end{figure*}

\paragraph{Reflectance Factorization}
In practice, the outgoing radiance is highly correlated with the surface's reflectance. As used in \cite{muller2020neural}, we decouple the texture and radiance by factorizing the reflectance out of the learning target. This allows the network to focus solely on the texture-independent details, alleviating the burden on the network by freeing it from the need to memorize texture information, as shown in Fig. \ref{fig:reflectance}.

\paragraph{Additional Inputs}
To enhance the model's ability to capture the radiance, we also provide auxiliary geometry features as additional inputs to the network. Specifically, we incorporate the outgoing direction $\omega_o$, as well as the albedo and surface normal at the hit point $\mathbf{x}$. The albedo is encoded using OneBlob encoding \cite{muller2019neural}, while the outgoing direction $\omega_o$ and surface normal are encoded with spherical harmonic basis functions.

\begin{table}[b]
\caption{Our approach vs. multi-resolution hash grid with varying hash table sizes for each level ($2^{17}-2^{19}$ features). Memory usage (MB) and performance (ms) in the encoding and MLP components in the static scene are reported. Our method’s vertex encoding consumes less memory than competing approaches, while also delivering a modest performance gain.}
\label{tab:static}
\begin{tabular}{cc|ccc}
\rowcolor[HTML]{FFFFFF} 
\multicolumn{2}{c|}{\cellcolor[HTML]{FFFFFF}Memory / Time} & Dining Room & Living Room & Veach Door   \\ \hline
\rowcolor[HTML]{F0F0F0} 
Hash17                      & Encoding                     & 13.9 / 32.2 & 13.9 / 32.2 & 13. 9 / 32.1 \\
\rowcolor[HTML]{FFFFFF} 
                            & MLP                          & 1.1 / 27.2  & 1.1 / 27.4  & 1.1 / 27.4   \\ \hline
\rowcolor[HTML]{F0F0F0} 
Hash18                      & Encoding                     & 33.9 / 32.1 & 33.9 / 32.1 & 33.9 / 32.1  \\
\rowcolor[HTML]{FFFFFF} 
                            & MLP                          & 1.1 / 27.7  & 1.1 / 27.5  & 1.1 / 27.4   \\ \hline
\rowcolor[HTML]{F0F0F0} 
Hash19                      & Encoding                     & 41.9 / 32.1 & 41.9 / 32.2 & 41.9 / 32.1  \\
\rowcolor[HTML]{FFFFFF} 
                            & MLP                          & 1.1 / 27.5  & 1.1 / 27.4  & 1.1 / 27.4   \\ \hline
\rowcolor[HTML]{F0F0F0} 
Ours                        & Encoding                     & 3.7 / 27.6  & 4.3 / 23.1  & 3.8 / 13.6   \\
\rowcolor[HTML]{FFFFFF} 
                            & MLP                          & 1.1 / 27.1  & 1.1 / 27.1  & 1.1 / 23.1   \\ \hline
\end{tabular}
\end{table}

\paragraph{Architecture \& Training}
For static and dynamic scenes, we assign 4 and 8 learnable parameters per vertex, respectively. The MLP has 4 hidden layers with 256 neurons each. Training uses Adam with learning rate \(1\times10^{-3}\), decayed by 0.33 every one-third of the total steps. The RHS sample count \(M\) (Eq.~\ref{eq:our_mc_residual}) starts at 32 and doubles every \(1/5\) of the training steps \cite{su2024dynamic}. LOD factors are updated three times early in training, limited to at most 50\% of the original vertex count (100\% for dynamic scenes) and clamped to \([0,30]\). The probability mixing factor \( \alpha \) in Eq.~\ref{eq:prob_mix} is fixed at 0.5.

\paragraph{Details of Hash Grid}  
The hash grid for static scenes is configured with eight resolution levels, with the coarsest resolution set to 4 and a scale factor of 2 applied between successive levels. Each voxel is assigned 8 learnable parameters per resolution level. To explore performance across different capacities, we set the hash map sizes to 17, 18, and 19 for comparison. For dynamic scenes, we adopt the same hyperparameter settings as specified in \citet{su2024dynamic}.
\section{Results}
\label{sec:result}

\begin{table}[b]
\caption{Our approach vs. DNR \cite{su2024dynamic} in the dynamic scenes. We report memory usage (MB) and performance (ms) statistics for the spatial encoding, extra encoding, and MLP components. Our method’s vertex encoding achieves substantially greater memory savings, even surpassing the results in the static scene, and delivers a modest performance improvement.}
\label{tab:dynamic}
\begin{tabular}{ccccc}
\rowcolor[HTML]{FFFFFF} 
\multicolumn{2}{c}{\cellcolor[HTML]{FFFFFF}Memory / Time}                    & Dining Room & Living Room & Veach Door  \\ \hline
\rowcolor[HTML]{F0F0F0} 
DNR & \multicolumn{1}{c|}{\cellcolor[HTML]{F0F0F0}Encoding} & 105 / 31.1  & 105 / 31.1  & 105 / 31.1  \\
\rowcolor[HTML]{FFFFFF} 
                     & \multicolumn{1}{c|}{\cellcolor[HTML]{FFFFFF}$f_\mathcal{XV}$}      & 14.0 / 16.0 & 10.0 / 10.2 & 12.0 / 13.4 \\
\rowcolor[HTML]{F0F0F0} 
                     & \multicolumn{1}{c|}{\cellcolor[HTML]{F0F0F0}$f_\mathcal{V}$}       & 0.3 / 1.8   & 0.3 / 1.6   & 0.3 / 1.7   \\
\rowcolor[HTML]{FFFFFF} 
                     & \multicolumn{1}{c|}{\cellcolor[HTML]{FFFFFF}MLP}      & 1.2 / 12.9  & 1.2 / 12.0  & 1.1 / 12.5  \\ \hline
\rowcolor[HTML]{F0F0F0} 
Ours                 & \multicolumn{1}{c|}{\cellcolor[HTML]{F0F0F0}Encoding} & 7.4 / 26.4  & 6.8 / 23.2  & 17.3 / 12.8 \\
\rowcolor[HTML]{FFFFFF} 
                     & \multicolumn{1}{c|}{\cellcolor[HTML]{FFFFFF}$f_\mathcal{XV}$}      & 14.0 / 16.1 & 10.0 / 10.2 & 12.0 / 13.3 \\
\rowcolor[HTML]{F0F0F0} 
                     & \multicolumn{1}{c|}{\cellcolor[HTML]{F0F0F0}$f_\mathcal{V}$}       & 0.3 / 1.8   & 0.3 / 1.6   & 0.3 / 1.7   \\
\rowcolor[HTML]{FFFFFF} 
                     & \multicolumn{1}{c|}{\cellcolor[HTML]{FFFFFF}MLP}      & 1.1 / 11.8  & 1.1 / 11.1  & 1.1 / 11.4 
\end{tabular}
\end{table}

\begin{figure*}[t]
  \centering
  \includegraphics[width=0.9\linewidth]{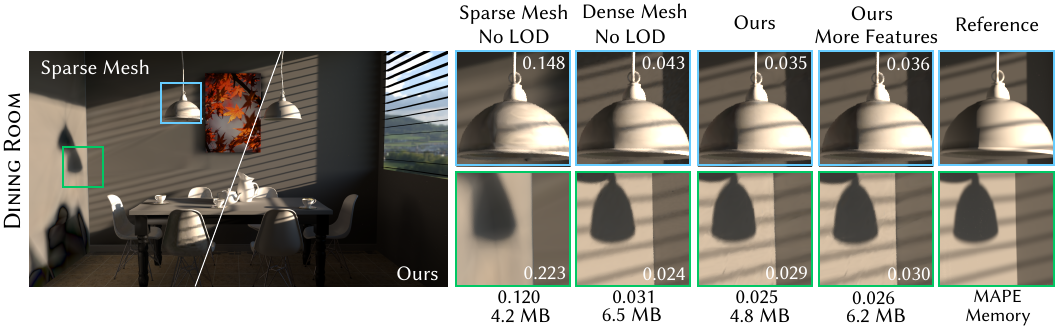}
  \vspace{-0.1in}
\caption{Varying strategies for capturing fine-grained details in the \textsc{Dining Room} scene. Sparse mesh with no LOD: Poorly represented high-frequency shadow details on the walls due to low vertex density. Dense mesh with no LOD: More detail but with higher memory consumption. Our method: Better results with lower memory consumption via adaptive LOD. Our method with more features: similar quality but with more memory usage.
}
  \label{fig:no_subdivide}
\end{figure*} 

\subsection{Static Scenes}

\begin{figure*}[t]
  \centering
  \includegraphics[width=\linewidth]{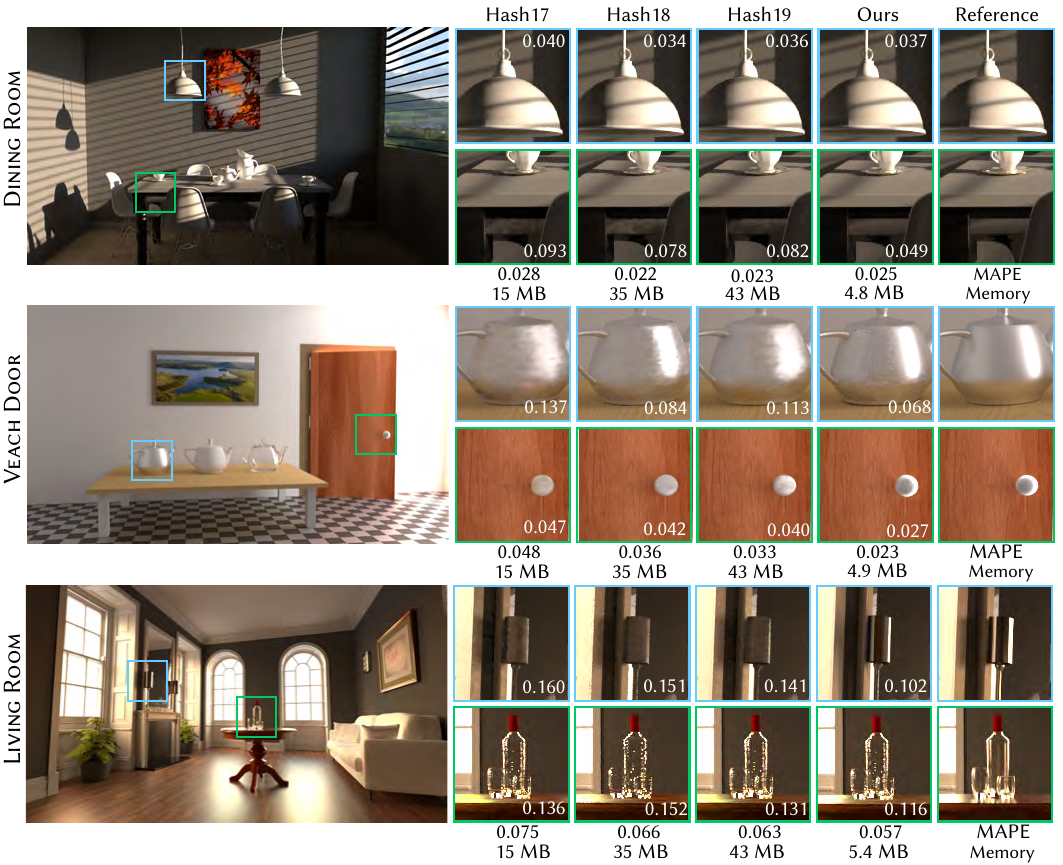}
\caption{Evaluating the effectiveness of neural vertex feature on Neural Radiosity \cite{hadadan2021neural}. We compare the visual quality of our method against multi-resolution hash encoding \cite{muller2022instant} with different hash table sizes ($2^{17}$-$2^{19}$ features for each level) in 3 static scenes. However, it tends to exhibit more blurriness on high-frequency details and grid-like artifacts as the feature capacity decreases. Our method achieves much less memory consumption at comparable visual quality by further culling the features from the empty scene spaces.}  
  \label{fig:static_result}
\end{figure*}

We first evaluate our method on the application of pre-trained surface-based neural caching in static scenes, namely Neural Radiosity \cite{hadadan2021neural}, where the positional network inputs could be backed with different encoding schemes, e.g., Fourier basis encoding \cite{mildenhall2021nerf}, and trainable parametric encoding \cite{muller2022instant, takikawa2023compact}. In this evaluation, we conduct qualitative experiments against a multi-resolution hash grid with varying hash table sizes for each level ($2^{17}-2^{19}$ features). Both methods are trained in $2-3$ hours. As shown in Fig. \ref{fig:teaser}, Fig. \ref{fig:static_result} and Tab. \ref{tab:static}, grid encoding tends to exhibit blurriness around fine details as the capacity of the hash table decreases. Moreover, the grid-like artifacts could be seen at the smooth transitions of shaded surfaces, which we attribute to the increased hashing collisions over the smaller hash table. On the other hand, our method achieves more accurate modeling of view-dependent surface shading by allocating spatial features in a more coherent manner with scene geometries, while explicitly accounting for the shading variations introduced by the geometric discontinuities along the vertex edges.

\subsection{Dynamic Scenes}

\begin{figure*}[h]
  \centering
  \includegraphics[width=\linewidth]{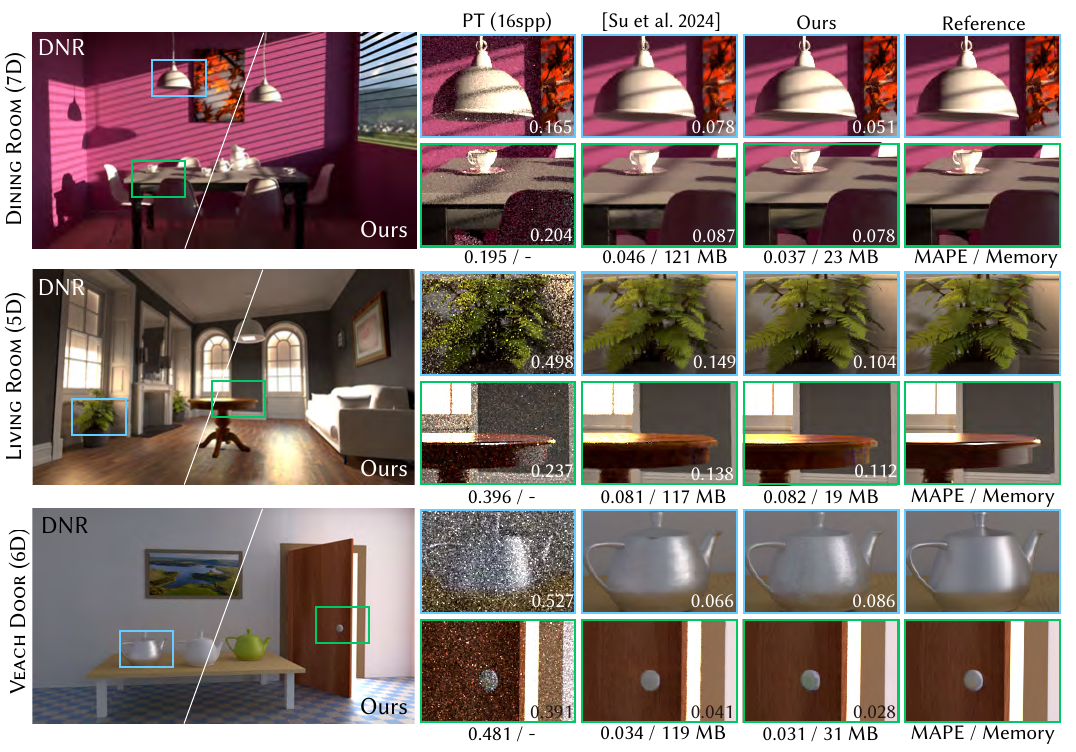}
  \caption{Evaluating the effectiveness in three dynamic scenes featuring various animated components. 
  Neural vertex features substitute the 3D hash grid of \citet{su2024dynamic} to encode purely spatial information, cutting memory and avoiding the grid-like artifacts that hashing’s random interpolation introduces in smooth regions while also relieving the network from handling hash collisions. The dynamic component remains represented by the original 2D feature planes and small MLPs. The surface-aligned locality of vertex features reproduces gradual shading transitions in both diffuse and specular areas. Please refer to the supplemental video for frame-by-frame comparisons.
  } 
  \label{fig:dynamic_result}
\end{figure*}

We then evaluate our method on the dynamic cases previously established by \citet{su2024dynamic}. We test the effectiveness of our encoding in dynamic scenes featuring different types of animated components and varying geometric complexities, including:
\begin{itemize}
\item The 7D \textsc{Dining Room} scene features rotational environment lighting, two dynamic ceiling lights, a wall with varying reflectance, and three animated objects.
\item The 5D \textsc{Living Room} scene includes window blinds for controlling incoming light, a ceiling light with adjustable intensity and height, and an animated table that moves along the $xy$ plane.
\item The 6D \textsc{Veach Door} scene consists of three animated meshes, an animated door with varying opening amplitude, and a floor surface with changing reflectance.
\end{itemize}
We train our model for $\sim$10 hours per scene and compare our approach with equal-time path tracing (PT) and the state-of-the-art neural rendering method for dynamic scenes, Dynamic Neural Radiosity (DNR) \cite{su2024dynamic}. As shown in Fig. \ref{fig:dynamic_result}, our method achieves visually comparable results with DNR but with far less memory consumption. Specifically, the original grid-based encoding used by \citet{su2024dynamic} tends to blur high-frequency details and exhibits grid-like artifacts. 
This is due to the fact that our formulation is object-oriented, which allows for efficient modeling of the transform-invariant features of animated meshes.
Additionally, our method achieves a roughly $5\times$ reduction of memory footprint with less inference overhead, with model statistics shown in Tab. \ref{tab:dynamic}. Additional demonstrations of the animated scene experiments can be inspected in the accompanied video.

\begin{figure*}[t]
  \centering
  \includegraphics[width=\linewidth]{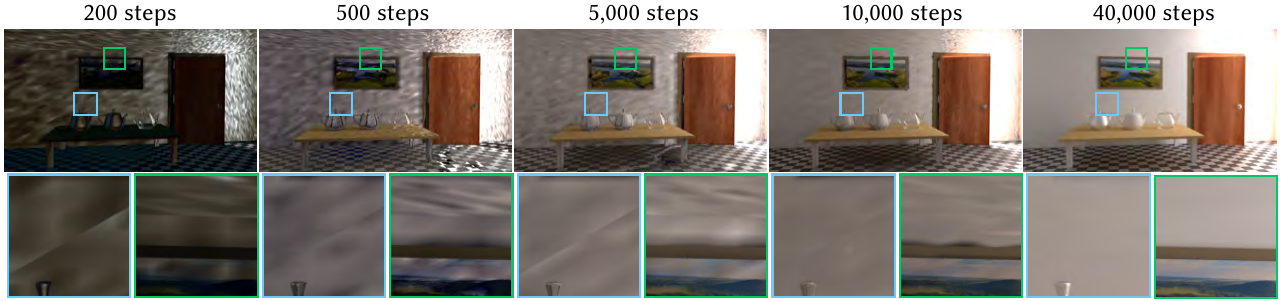}
  \caption{Discontinuities at triangle boundaries caused by different LOD levels gradually vanish as training progresses. Our neural network–based approach effectively smooths these discontinuities, yielding a seamless feature representation across triangle boundaries, even when different LOD levels are used.}
  \label{fig:continuous}
\end{figure*}

\subsection{Evaluation}

\begin{figure*}[t]
  \centering
  \includegraphics[width=0.95\linewidth]{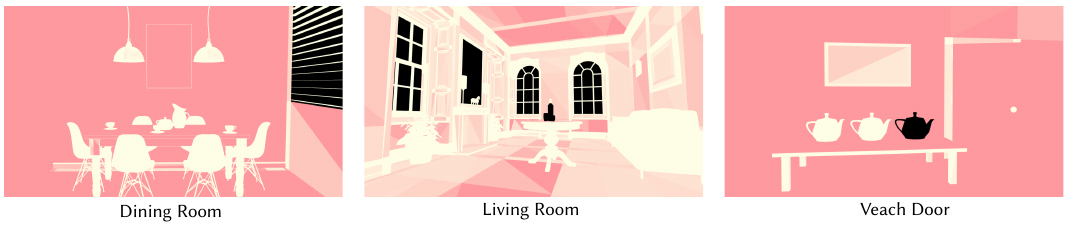}
\caption{Visualization of the learned LOD factors (relatively higher subdivision levels are shown in warmer colors). Refinement is concentrated in regions with low initial mesh density, such as floors and walls, where finer detail is required. This highlights the effectiveness of our method in adaptively allocating virtual vertices to under-represented areas, while avoiding unnecessary refinement in well-resolved regions.}  
  \label{fig:subdivide_level}
\end{figure*}

\paragraph{Visualization of Adaptive LOD}
We visualize the adaptive LOD results in the static scene rendering experiment, as shown in Fig.~\ref{fig:subdivide_level}. The results show that most multi-resolution refinement occurs in regions with low initial mesh density, such as the floor and walls, where additional detail is needed for accurate reconstruction. This demonstrates that our method effectively identifies under-represented areas and adaptively allocates virtual vertices, enabling improved feature representation without unnecessary refinement in well-resolved regions.

\paragraph{Ablation Study on Multi-resolution}
We evaluate the effectiveness of our multi-resolution strategy through a comparison of various mesh resolutions with/without LOD. As shown in Fig.~\ref{fig:no_subdivide}, the \textsc{Dining Room} scene contains high-frequency shadow details on the walls, which are poorly represented due to the low vertex density in the original mesh. Without LOD, these regions lack sufficient feature support, resulting in noticeable loss of detail. Increasing mesh density through manual subdivision can improve detail, but is resource-intensive. In contrast, our method leverages an adaptive multi-resolution strategy to achieve superior detail preservation using the original sparse mesh as input, while requiring less memory. This highlights the importance of adaptive LOD in capturing fine-grained appearance in under-resolved areas.

\paragraph{Feature Continuity Across Triangle Boundaries}
When adjacent triangles use different LOD levels, features across their shared edges may introduce discontinuities. As shown in Fig. \ref{fig:continuous}, rendering results at different training steps reveal this effect. At early stages (steps 200 and 500), visible seams appear along the boundaries. However, as training progresses (steps 5,000 and 10,000), the seams become less noticeable. By the final step (40,000), the discontinuity is no longer perceptible. This demonstrates that, given sufficient training, the network effectively smooths out the discontinuities, producing a seamless feature representation across triangle boundaries, even with different LOD levels.

\begin{figure}[h]
  \centering
  \includegraphics[width=0.9\linewidth]{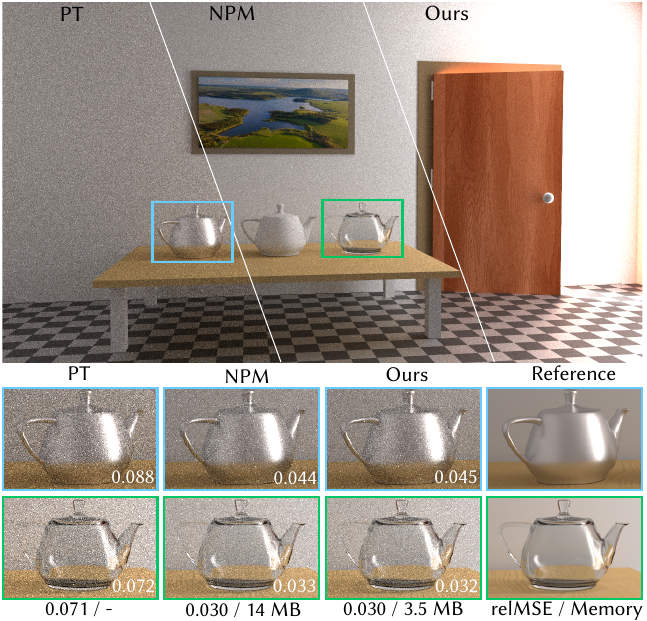}
  \caption{We validate our approach on the neural path guiding techniques \cite{dong2023neural}. Our method achieves comparable quality with the hashed grid approaches while using much less memory footprint. }
  \label{fig:path_guiding_result}
\end{figure}

\paragraph{Neural Path Guiding}
We also validate the effectiveness of our method on another family of techniques leveraging neural caches to facilitate rendering, namely neural path guiding \cite{dong2023neural, huang2024online}. This technique re-parameterizes the network outputs to analytic models (e.g., mixtures of spherical Gaussians) for efficient guided sampling and variance reduction. 
In this evaluation, we implement their method in Mitsuba3 renderer using PyTorch and apply our technique to it. Following the configuration of \citet{dong2023neural}, we let the network predict mixtures of normalized spherical Gaussians with 8 lobes. The learned mixtures are trained online and used to do multiple importance sampling (MIS) \cite{veach1995optimally} with original BSDF sampling at each shading point in unidirectional path tracing. Since the path guiding task involves only a single-view rendering and requires less training time, we use a pre-subdivided mesh to provide sufficient feature resolution. As shown in Fig. \ref{fig:path_guiding_result}, our method achieves comparable quality with the hashed grid approaches while using much less memory footprint.

\section{Conclusion, Limitations, and Future Work}
\label{sec:conclusion}

In this work, we propose a trainable latent encoding for efficient neural rendering. By encoding spatial information directly at the vertices of the mesh, we achieved more compactness by exploiting the geometric priors, thereby reducing memory overhead compared to traditional 3D-grid-based encoding. Our method captures important surface details while maintaining computational efficiency. We boost low-density meshes with a virtual multi-resolution scheme that refines features where training loss is high, enabling recovery of high-frequency lighting in undersampled areas. Our method achieves significant memory savings and comparable or better image quality compared to state-of-the-art methods. Furthermore, our vertex feature formulation has the potential to perform various neural rendering tasks.

Despite the effectiveness of our method, there are still some limitations. Since our approach encodes features directly at mesh vertices, it is inherently restricted to surface-based representations and cannot be directly applied to volumetric data or scenes without explicit surface geometry. Additionally, our method focuses on subdividing features at the vertices without simplifying the underlying mesh structure. This design choice was made to preserve geometric detail, as mesh simplification may result in a loss of details. However, this could be explored as a future improvement for more efficient representations while retaining high expressiveness.

In dynamic scenes from the video, we still observe subtle artifacts and frame-to-frame inconsistencies, whereas these issues are absent in static scenes.
This is due to the dynamic scene’s high dimensionality and the reliance on grid-based encodings ($f_\mathcal{XV}$ and $f_\mathcal{V}$) for modeling the dynamic components. More broadly, using a shallow MLP to model such components in a high-dimensional space can induce temporal instabilities (the same issue with DNR). In particular, when moving objects cast shadows beneath meshes, the resulting regions contain high-frequency radiance that the network struggles to capture. This will be one of our future works to tackle this issue.

Moreover, we also plan to explore hybrid representations that combine surface-based encoding with sparse volumetric features to support more diverse scene types. Furthermore, we aim to investigate methods for mesh simplification in conjunction with feature refinement, allowing for more efficient representation while maintaining high expressiveness. We also note that our method could be straightforwardly applied to alternative geometry primitives, such as quadrilaterals, and we leave the validation of its practicality and effectiveness for future exploration.



\bibliographystyle{ACM-Reference-Format}
\bibliography{references}

\appendix

\end{document}